\documentclass[sn-mathphys-ay, Namedate, iicol]{sn-jnl}
\usepackage{graphicx}    
\usepackage{amsmath}     
\usepackage{amssymb}     
\usepackage{amsfonts}    
\usepackage{xspace}
\usepackage{amsthm}%
\usepackage{mathrsfs}%
\usepackage[title]{appendix}%
\usepackage{xcolor}%
\usepackage{textcomp}%
\usepackage{manyfoot}%
\usepackage{booktabs}%
\usepackage{listings}%
\newcommand{\Msun}{\,$M_{\odot}$\xspace}
\newcommand{\Rsun}{\,$R_{\odot}$\xspace}

\newcommand{\kms}{\,km\,s$^{-1}$\xspace}

\newcommand{\gcm}{\,g\,cm$^{-1}$\xspace}
\newcommand{\gcmq}{\,g\,cm$^{-3}$\xspace}

\newcommand{\Ha}{H$\alpha$\xspace}
\newcommand{\Hb}{H$\beta$\xspace}
\newcommand{\Hg}{H$\gamma$\xspace}

\newcommand{\HeII}{He\,{\sc ii}\xspace}
\newcommand{\CIII}{C\,{\sc iii}\xspace}

\newcommand{\NIII}{N\,{\sc iii}\xspace}

\newcommand{\FeII}{Fe\,{\sc ii}\xspace}

\newcommand{\A}{\,\AA\xspace}
\newcommand{\apj}{Astrophys. J. } 
\newcommand{\apjl}{Astrophys. J. Lett. } 
\newcommand{\apss}{Astrophys. Space Sci. } 
\newcommand{\aap}{Astron. Astrophys. } 
\newcommand{\mnras}{Mon. Not. R. Astron. Soc. } 
\newcommand{\nat}{Nature } 
\newcommand{\pasj}{Publ. Astron. Soc. Jpn } 
\begin{document}
\title[Type IIP SN~2023ixf] 
{Constraining hydrodynamic model of nearby type IIP SN~2023ixf}
 \author*[1,2]{\fnm{V. P.} \sur{Utrobin}}\email{utrobin@itep.ru}

\author[2]{\fnm{N. N.} \sur{Chugai}}\email{nchugai@inasan.ru}


\affil*[1]{\orgname{NRC ``Kurchatov Institute''},
   \orgaddress{\street{acad. Kurchatov Square 1}, \city{Moscow},
   \postcode{123182}, \country{Russia}}}

\affil[2]{\orgname{Institute of Astronomy, Russian Academy of Sciences},
   \orgaddress{\street{Pyatnitskaya St. 48}, \city{Moscow},
   \postcode{119017}, \country{Russia}}}

\abstract{%
Despite proximity of SN~2023ixf and a wealth of observational data,
   the released hydrodynamic models leave too broad range of the derived
   explosion energy and the ejected mass.
We revisit the hydrodynamic modeling based on a broader set of observables
   than have been previously used.
Among those of top priority is the early maximum ejecta velocity that is crucial
   in removing parameter degeneracy.
The inferred parameters of SN~2023ixf are the explosion energy of
   $2.8\times10^{51}$\,erg, ejecta mass of 13.2\Msun, presupernova radius of
   1540\Rsun, and $^{56}$Ni mass of 0.07\Msun.
The circumstellar matter is composed by the dense circumstellar shell with
   the mass of 0.01\Msun and radius of $5\times10^{14}$\,cm, as well as
   the external rarefied wind.
Both circumstellar components are consistent with the early \Ha broad wings
   caused by the Thomson scattering and the intrinsic column density provided
   by X-ray data.
Based on the radiation hydrodynamics we, for the first time, simulate
   the SN~2023ixf phenomenon from the explosion to the emergence of
   the hard X-ray radiation.
}

\keywords{hydrodynamics -- methods: numerical -- supernovae: general --
   supernovae: individual: SN~2023ixf
}

\maketitle

\section{Introduction}
\label{sec:intro}
Type IIP supernova (SN) 2023ixf discovered by \cite{Itagaki_2023} in nearby
   galaxy M101 at the distance of $6.19-6.74$\,Mpc (NED Database) immediately
   became a target for multi-band observations.
This provides us an opportunity to study the explosion of a red supergiant (RSG)
   in more detail than ever before and to clarify unresolved issues.

At the first place is a problem of the reliability of key SN parameters
   (the explosion energy, pre-SN mass, and radius) recovered via
   a hydrodynamic modeling.
Contrary to all expectations, however, the released hydrodynamic models of
   SN~2023ixf \citep{Bersten_2024, Moriya_2024, Forde_2025, Kozyreva_2025,
   Hsu_2025, Vinko_2025, Laplace_2025} leave us with a broad choice of
   the inferred explosion energy: $(0.5-3)\times10^{51}$\,erg.
This situation is highly alarming, since it undermines the very idea that
   we could infer reliable parameters of SN~2023ixf.

Fortunately, there is a way to significantly reduce the uncertainty of
   hydrodynamic model --- to use a broader set of available observational data
   with an emphasis on key observables.
The first, and the most important among those, is the maximum ejecta velocity
   at the early stage --- an indispensable observable for removing a parameter
   degeneracy.
This possibility has been overlooked in the published models of SN~2023ixf.
Meanwhile, the shallow Balmer line absorptions detected in early spectra
   \citep{Zheng_2025} can provide us with early expansion velocities.

The second important observable is the velocity extent of the $^{56}$Ni ejecta
   that can be recovered from the [Co\,{\sc ii}] 10.521\,$\mu$m emission line
   in JWST spectrum on day 252 \citep{Medler_2025}.
The end part of the light curve plateau depends on the $^{56}$Ni extent and
   knowing this parameter permits us to more reliably fix the ejecta mass.
The third constraint is related to the circumstellar (CS) shell density and
   its extent; both are imprinted in the early \Ha profile and the column
   density inferred from X-ray data \citep{Grefenstette_2023, Nayana_2025}.
The early emergence of the X-ray flux provides also an additional test for
   the hydrodynamic model.

To avoid an ambiguity related to the term ``CS shell'', it should be emphasized
   that some hydrodynamic models \citep{Bersten_2024, Moriya_2024, Laplace_2025}
   include the massive CS shell ($\sim$0.4$-0.5$\Msun) attached to the RSG
   in order to describe an initial ($t < 20$ days) luminosity peak of SN~2023ixf.
Such a massive CS shell is not needed in our model.
We consider instead the explosion of an extended RSG model with the attached
   low-mass CS shell indicated by early ionization flash spectra
   \citep{Bostroem_2023}.
The expected mass of this shell based on the case of SN~2013fs is
   $\sim$0.003\Msun \citep{Yaron_2017} and could be as high as 0.006\Msun
   based on SN~2024bch \citep{UC_2025}.
Despite the low mass, the CS shell is an essential ingredient of the model
   for the early hard X-rays.
The point is that the viscous jump in the forward shock depends on
   the radiative acceleration of the preshock gas and the shock wave cooling,
   both being affected by the CS density.

The primary goal of the present paper is the hydrodynamic model of SN~2023ixf
   based on the extended set of observational data including those that
   have not been used so far to constrain the model.
We first consider relevant observational data and obtain some useful numbers
   (Sect.~\ref{sec:guide}).
Using available observational constraints, we then produce the appropriate
   hydrodynamic model (Sect.~\ref{sec:hdm}) and explore the early emergence
   of hard X-rays based on the radiation hydrodynamic modeling of
   the CS interaction (Sect.~\ref{sec:xrays}).

Below we adopt the distance $D = 6.71$\,Mpc and the reddening
   $E(B-V) = 0.0387$\,mag \citep{Hsu_2025}.
The explosion date is set to be MJD 60082.641, which we recover from
   the model fit to the earliest $r$ magnitudes \citep{Li_2024}.

\section{Observational guidelines}
\label{sec:guide}
The luminosity and photospheric velocity at the light curve \emph{plateau}
   poorly constrain hydrodynamic model of SN~IIP \citep{Utrobin_2007,
   Goldberg_2019, UC_2019}; this fact is sometimes referred to as a ``parameter
   degeneracy'' \citep{Goldberg_2019}.
Below we describe the key observables that remove the parameter degeneracy
   and permit to efficiently constrain the hydrodynamic model of SN~2023ixf.

\subsection{Early spectral evolution}
\label{sec:three}
The early evolution of SN~2023ixf spectra \citep{Bostroem_2023, Bostroem_2024})
   proceeds through three stages that reflect key physics of early SN evolution
   \citep{UC_2025} and manifests itself in the behavior of the hydrodynamic
   model.

Stage A ($t < t_1 \sim 5.5$ days).
The spectrum is a smooth continuum with emission lines showing a specific profile
   --- narrow core with broad wings.
The wings are related to the multiple Thomson scattering, while the SN expands
   in a dense CS shell with the significant Thomson optical depth.
The CS shell is swept up into a cold dense shell (CDS) by the end of this stage
   ($t = t_1$).

Stage B ($t_1 < t < t_2 \approx 18$ days).
The overall spectrum is a smooth continuum devoid of both pronounced narrow and
   broad lines.
The ejecta bounded by the opaque CDS runs with a constant velocity throughout
   the outer rarefied RSG wind.
The photosphere resides at the opaque CDS that conceals the unshocked SN ejecta.

Stage C ($t > t_2$).
Broad lines with the P Cygni profile emerge in the spectrum indicating that
   the CDS becomes transparent thus revealing the unshocked SN ejecta with
   the receding photosphere.

\subsection{Early maximum ejecta velocity}
\label{sec:vel}
%
\begin{figure}
   \includegraphics[width=\columnwidth, clip, trim=40 239  44 121]{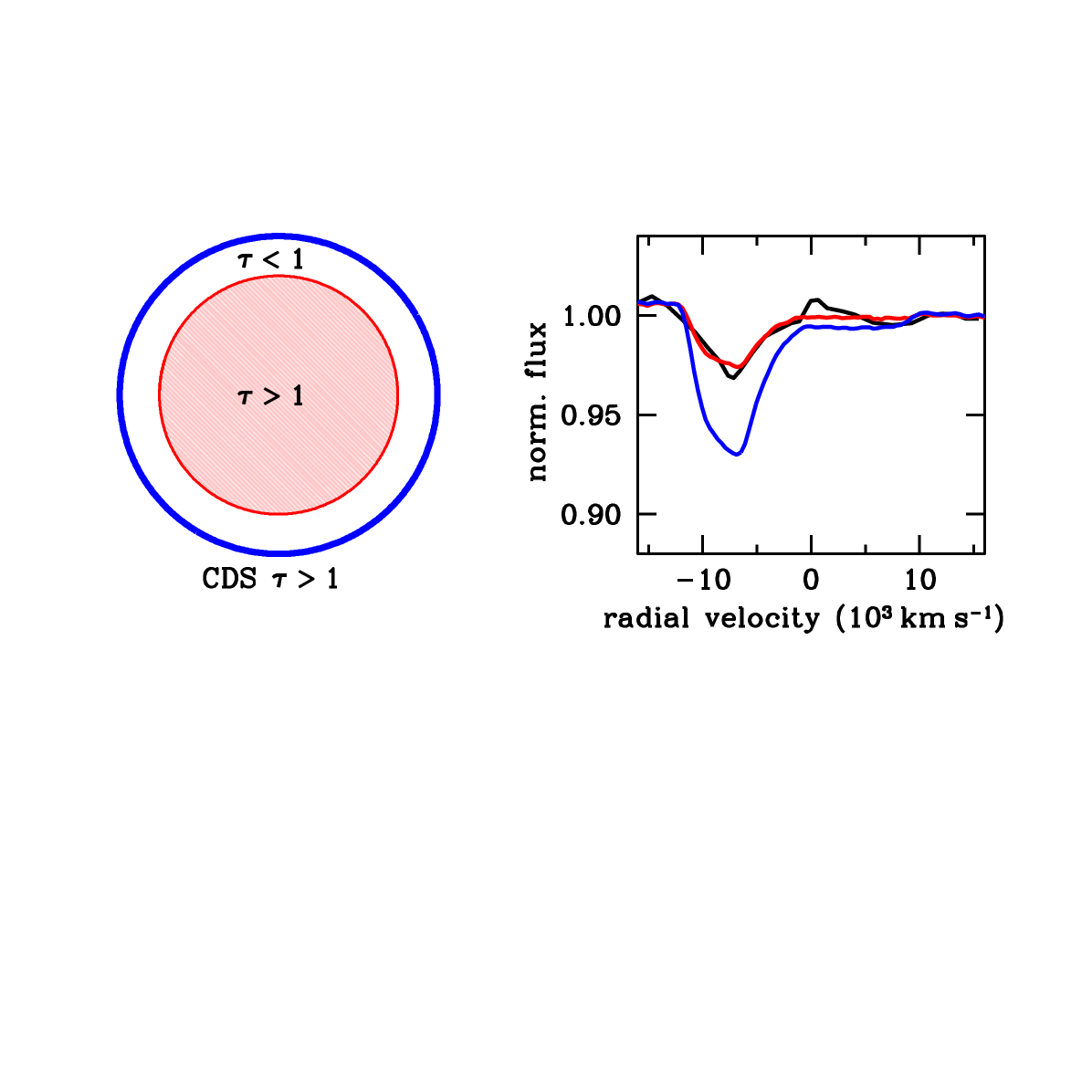}
   \caption{%
   Left: schematic visualization for the origin of shallow Balmer line
      absorptions in SN~2023ixf at the age of $8 - 14$ days.
   Absorption lines originate in the layer between the opaque core
      (\emph{pale-red disk\/}) and the opaque boundary CDS
      (\emph{blue circle\/}).
   Right: the model \Hg profile for the CDS optical depth of $\tau = 4$
      (\emph{red line\/}) overplotted on the observed \Hg absorption
      on day +9.4 (\emph{black line\/}).
   The model for $\tau = 3$ demonstrates the strong effect of the CDS
      optical depth on the depth of the absorption line.
   }
   \label{fig:shabal}
\end{figure}
Early expansion velocity of the ejecta in most SNe~IIP can be measured from
   the broad absorption of Balmer lines.
This, however. cannot be done for SNe~IIP with an opaque CDS that fully conceals
   the ejecta.
Surprisingly, the high-quality spectra of SN~2023ixf between days 8 and 14
   \citep{Zheng_2025} show broad Balmer line absorptions, though with a very
   low depth ($\sim$2$-3\%$ of the continuum).

We identify these absorptions with the outermost ejecta.
The \Ha line with the conservative scattering produces less pronounced
   absorption compared to that of the \Hb and \Hg lines with low scattering
   albedo.
Note that a similar depth of \Hb and \Hg indicates that both absorptions
   are saturated, i.e., the Sobolev optical depth of both lines
   $\tau_{\mbox{\tiny S}} \gg 1$.
This raises a question: why the absorptions of optically thick lines are
   extremely shallow?

The explanation is based on the structure of the outer ejecta at this stage
   (Fig.~\ref{fig:shabal}) with three relevant components: (i) the opaque ejecta
   core --- source of the SN luminosity; (ii) the opaque CDS; (iii) the ejecta
   layer between the core and the CDS, which is optically thin in the continuum.

The core continuum luminosity $L_{\nu}$ modified by absorption lines enters
   the CDS as $L_{\nu}r_{\nu}$, where  $r_{\nu}$ is the line (e.g., \Hg)
   residual intensity.
The unabsorbed luminosity transmitted through the CDS is
   $L_1 = L_{\nu}r_{\nu}e^{-\tau}$, where $\tau$ is the CDS continuum optical
   depth.
The luminosity absorbed and reprocessed by the CDS is
   $L_2 = L_{\nu}(1 - e^{-\tau})$.
Superposition $L = L_1 + L_2$ results in the observed luminosity
\begin{equation}
L =	r_{\nu}^{out}L_{\nu} = r_{\nu}L_{\nu}e^{-\tau} +  L_{\nu}(1 - e^{-\tau})\,.
	\label{eq:resid}
\end{equation}  
Here we omit a continuum correction related to the line absorption
   $\mathcal{O}[(v_{sn}/c)L_{\nu}]$.
The equation (\ref{eq:resid}) implies the observed line relative depth
   ($A_{\nu} = 1 - r_{\nu}$) to be $A_{\nu}^{out} = A_{\nu}e^{-\tau}$,
   where $A_{\nu} \approx 1$ for saturated absorptions.
The observed line depth of $A_{\nu}^{out}\sim 0.02$ then is produced,
   if the CDS optical depth is $\tau \approx 4$.

The above analytical consideration is supported by the Monte Carlo simulation.
We consider the \Hg line with the scattering albedo determined by the ratio of
   spontaneous radiative rates $\omega = A_{52}/(A_{52}+A_{53}+A_{54}) = 0.33$.
The Sobolev optical depth is set to be
   $\tau_{\mbox{\tiny S}} = \tau_{\mbox{\tiny S,c}}(v_c/v)^3$,
   where $\tau_{\mbox{\tiny S,c}}$ is the line optical depth at the core
   boundary.
The model (Fig.~\ref{fig:shabal}) satisfactory fits the observed \Hg
   on day +9.4\footnote{%
The sign plus henceforth denotes the time relative to the phase zero adopted
   by observers, whereas the time without plus is counted from the explosion
   date of the present paper.}
   \citep{Zheng_2025} for the CDS optical depth $\tau_{cds} = 4$,
   in accord with the analytical estimate.
The optimal boundary velocity of the ejecta is $v_{sn} = 12000\pm500$\kms,
   $v_c = 6500$\kms, and $\tau_{\mbox{\tiny S,c}} = 8$.
The observed absorption depth is insensitive to the $\tau_{\mbox{\tiny S,c}}$
   value, but rather sensitive to the CDS optical depth, which is demonstrated
   by the case of $\tau_{cds} = 3$ (Fig.~\ref{fig:shabal}).

The CDS velocity is expected to be close to the maximum ejecta velocity,
   because at this stage the CDS expands in the rarefied wind that exerts
   a negligible dynamic pressure.
This conjecture is confirmed by the constant velocity of blue absorption edge
   of the Balmer lines $v_{sn} \approx 12000$\kms between days +8.4 and +12.4
   \citep{Zheng_2025}.
Therefore, the expected CDS velocity at the given stage is
   $v_{cds} \approx 12000$\kms, which should be manifested in a constant
   photospheric velocity.

\subsection{$^{56}$Ni mixing}  
\label{sec:nimix}
The velocity extent of the $^{56}$Ni ejecta is an important parameter that
   controls a contribution of the radioactive heating to the luminosity
   at the end of the light curve plateau and at the transition to
   the radioactive tail.
Usually, the outer $^{56}$Ni velocity is not available from observations of
   SNe~IIP, except for SN~1987A.
Fortunately, the JWST spectrum of SN~2023ixf on day +252.67
   \citep{Medler_2025} shows a high-quality [Co\,{\sc ii}] 10.521\,$\mu$m
   emission line with the blue width of 4000\kms at zero intensity.
We attribute this value to the outer velocity of the $^{56}$Ni ejecta.
Knowing the $^{56}$Ni velocity extent strengthens the reliability of
   the ejecta mass inference.

\subsection{CS shell density} 
\label{sec:csm}
%
\begin{figure}
   \includegraphics[width=\columnwidth, clip, trim=9 240  55 115]{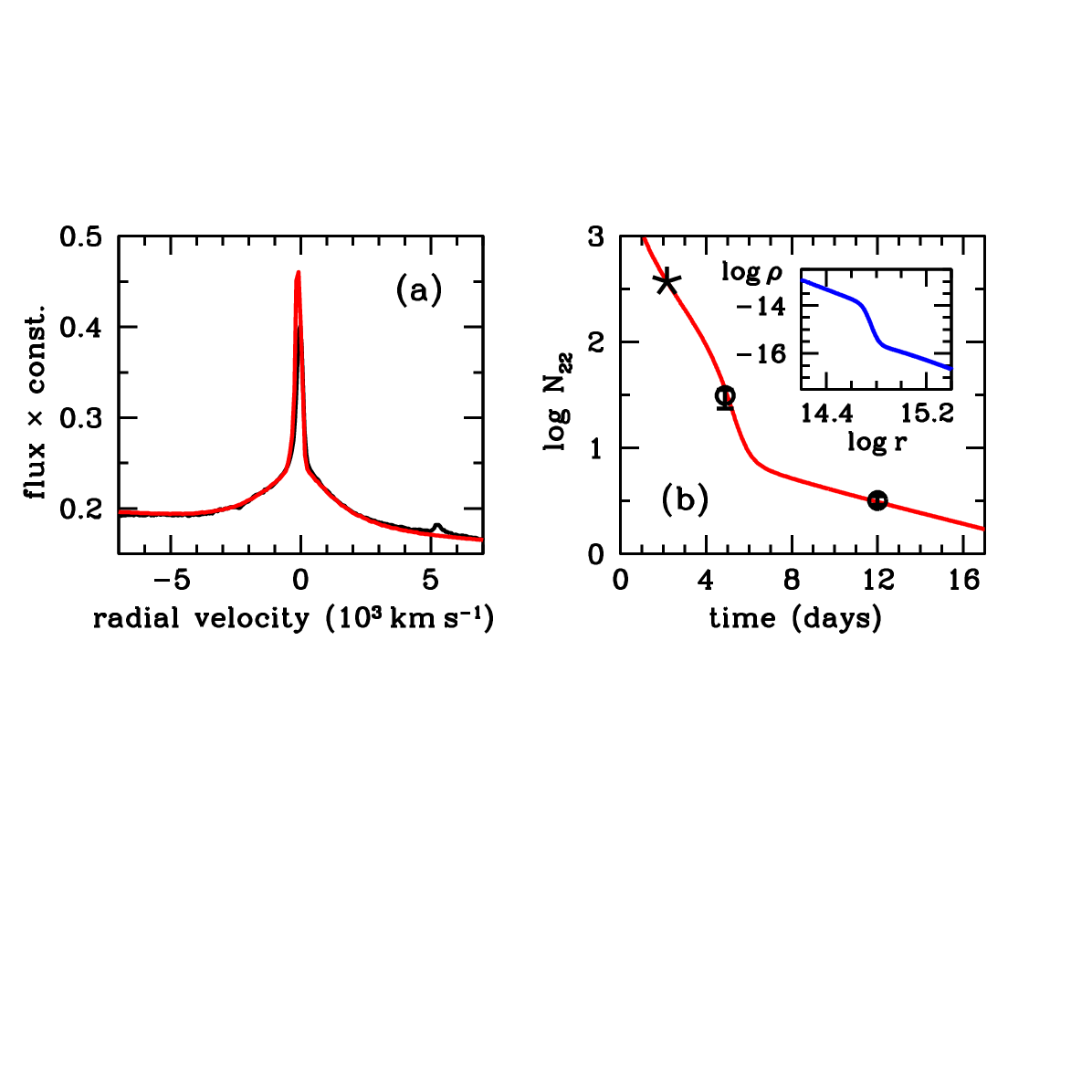}
   \caption{%
   Observational effects of the CSM.
   Panel (a): the model \Ha profile (\emph{red line\/}) overlaid on
      the observed \Ha on day +1.67 (\emph{black line\/})
      \citep{Bostroem_2023}.
   Panel (b): the evolution of the column density (\emph{red line\/}) that
      is suggested by the \Ha profile and the fit to the $N_{\mbox{\tiny H}}$
      values recovered from X-ray spectra for two epochs \citep{Nayana_2025}.
   The \emph{star} symbol shows the $N_{\mbox{\tiny H}}$ values suggested by
      the Thomson optical depth.
   Inset shows the density of the CS shell suggested by the \Ha profile and
      the density of the external wind obtained from the X-ray column density.
   }
   \label{fig:csha}
\end{figure}
Broad \Ha wings at the stage A are formed due to the Thomson scattering with
   some contribution of the CS expansion velocities.
It is highly important to rely on the earliest spectra in order to minimize
   the effect of the preshock velocities of the accelerated CS gas.
For analysis we use the MMT spectrum \citep{Bostroem_2023} on day +1.67
   or on day 1.78 from the SN explosion.

Our model suggests that the \Ha line at this stage forms outside the opaque CDS
   ($r = r_0$, see below) with the adopted albedo of a diffuse reflection
   $A =  0.16$.
This value is the average over cosine of the incident angle assuming
   a single scattering albedo of 0.5 \citep{Sobolev_1975}.
We find that the result is not sensitive to this parameter.
Given the CDS velocity $v_{cds} = 12000$\kms, the CS shell at the moment
   $t_0 = 1.78$ days occupies the range $r_0 < r < r_1$ with
   $r_0 = v_{cds}t_0 = 1.8\times10^{14}$\,cm and
   $r_1 = v_{cds}t_1 = 5.7\times10^{14}$\,cm (where $t_1 =5.5$ days).
The dense internal wind ($r < r_1$) follows the law $\rho \propto r^{-2}$
   with a smooth transition at $r \approx r_1$ to the rarefied external wind.
The density of this internal wind is constrained by the Thomson optical depth,
   whereas the external wind density is constrained by the column density
   inferred from X-ray data.
The CS shell gas can be radiatively accelerated with the velocity profile
   $u = u_0(r_0/r)^2$, where $u_0$ is a free parameter.
The electron temperature of the CS shell is assumed to be $3\times10^4$\,K,
   in a qualitative accordance with the presence of high ionization emission
   lines of \HeII, \CIII, and \NIII \citep{Bostroem_2023}.
     
The primary free parameters of the model are the Thomson optical depth
   $\tau_{\mbox{\tiny T}}$ and the preshock velocity of the accelerated CS gas
   $u_0$.
The \Ha emissivity is assumed to be determined by the recombination rate
   $\epsilon \propto n_e n_{\mbox{\tiny H}}$, which may be not the case
   in reality.
Indeed, the observed spectrum on day +1.67 shows the \Ha/\Hb flux ratio of
   $\approx$1.5, significantly lower than the value of 2.7 expected for
   the recombination case B \citep{Osterbrock_2006}.

The \Ha profile is calculated using the Monte Carlo technique in
   the approximation of the instant radiation escape.
Since the \Ha recombination emissivity is a crude approximation, we use
   the normalized model to fit the observed \Ha emission line
   (Fig.~\ref{fig:csha}a).
The apparent excess of the model narrow \Ha component is related to
   the approximation of the recombination emissivity.
Indeed, the SN radiation can depopulate the upper hydrogen levels thus
   suppressing recombination emissivity.

The inferred parameter values are $\tau_{\mbox{\tiny T}} = 3$ and $u_0 = 570$\kms.
The preshock velocity turns out small and does not affect the recovered Thomson
   optical depth.
The diffusion time in the CS shell 
   $t_{dif} \approx (r_0/c)\tau_{\mbox{\tiny T}} \approx 10^4$\,s
   is small compared to the considered age, which justifies the approximation
   of the instant radiation escape.

The inset of Fig.~\ref{fig:csha}b shows the model density distribution in
   the CS shell and the external wind; the latter is constrained by the column
   density on day +11.46 recovered from the X-ray spectrum
   \citep{Grefenstette_2023, Nayana_2025}).
For the found density distribution the evolution of the column density outside
   the CDS is consistent with the both values inferred from X-ray data.

To summarize, the CSM is composed of the CS shell (internal dense wind) and
   the external rarefied wind.
The CS shell with the mass of $\approx 0.01$\Msun lies inside the radius of
   $5\times10^{14}$\Rsun and is characterized by the wind density parameter
   $w_{int} = 4\pi r^2\rho = 3.7\times10^{16}$\gcm.
The density parameter of the rarefied external wind is
   $w_{ext} = 1.7\times10^{15}$\gcm.

\section{Hydrodynamic modeling}
\label{sec:hdm}
\subsection{Supernova model}
\label{sec:snmod}
%
\begin{figure}
   \includegraphics[width=\columnwidth, clip, trim=0 239  54 121]{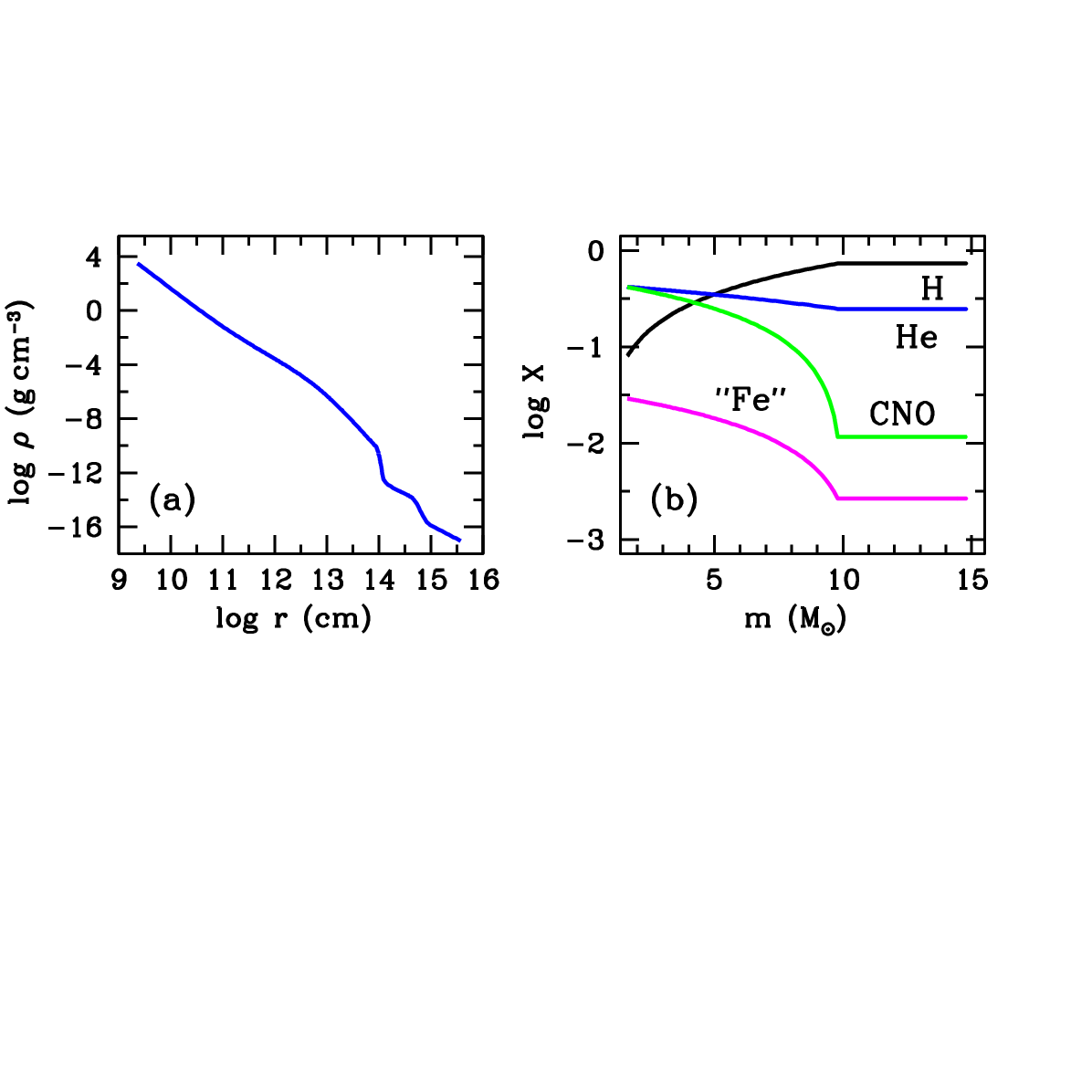}
   \caption{%
   The structure of the pre-SN model.
   Panel (a): the density distribution as a function of radius.
   At the radii $r > 10^{14}$\,cm the density refers to the CSM.
   Panel (b): the chemical composition as a function of mass coordinate.
   Mass fraction of hydrogen (\emph{black line\/}), helium
      (\emph{blue line\/}), CNO elements (\emph{green line\/}),
      and Fe-peak elements excluding radioactive $^{56}$Ni
      (\emph{magenta line\/}) in the ejected envelope.
   The central core of 1.6\Msun is omitted.
   }
   \label{fig:presn}
\end{figure}
The radiation hydrodynamic modeling of SN~2023ixf is performed using
   the Lagrangian code {\sc CRAB} \citep{Utrobin_2004, Utrobin_2007}.
The recent code modification implements an artificial mixing of the outermost
   layers \citep{Blinnikov_1998} to describe a boundary thin dense shell
   formed during the shock  breakout (SBO) \citep{GIN_1971, Chevalier_1976}.

The explosion is initiated by a supersonic piston at the boundary with
   the 1.6\Msun core that collapses into the neutron star.
The pre-SN is the hydrostatic RSG model based on the chemical composition of
   a ZAMS 17\Msun star before the collapse \citep{Woosley_2002}.
However the distribution of the density and the composition in the pre-SN is
   significantly modified in order to meet the observational data.

The need for the modification of the evolutionary model is imposed by the fact
   that the one-dimensional SN explosion of any evolutionary RSG star
   cannot reproduce the light curve and the expansion velocities of SN~IIP
   \citep{UC_2008, Utrobin_2017}.
The physical background for this fact is related to the three-dimensional (3D)
   mixing caused by the RSG explosion and 3D-effects in the extended convective
   hydrogen envelope at the final evolution stage \citep{Goldberg_2022}.

One of the target for the pre-SN modification is the density gradient of
   the hydrogen-rich envelope that should be steeper compared to
   the evolutionary model.
The distribution of the density and the composition of pre-SN is presented
   in Fig.~\ref{fig:presn}.
The model includes also a dense CS shell and an external RSG wind;
   both components are constrained by the early spectral evolution,
   the \Ha model, and the CSM column density recovered from the X-ray data
   (Sect.~\ref{sec:csm}).

\begin{figure*}
   \includegraphics[width=\textwidth, clip, trim=3 196  50 116]{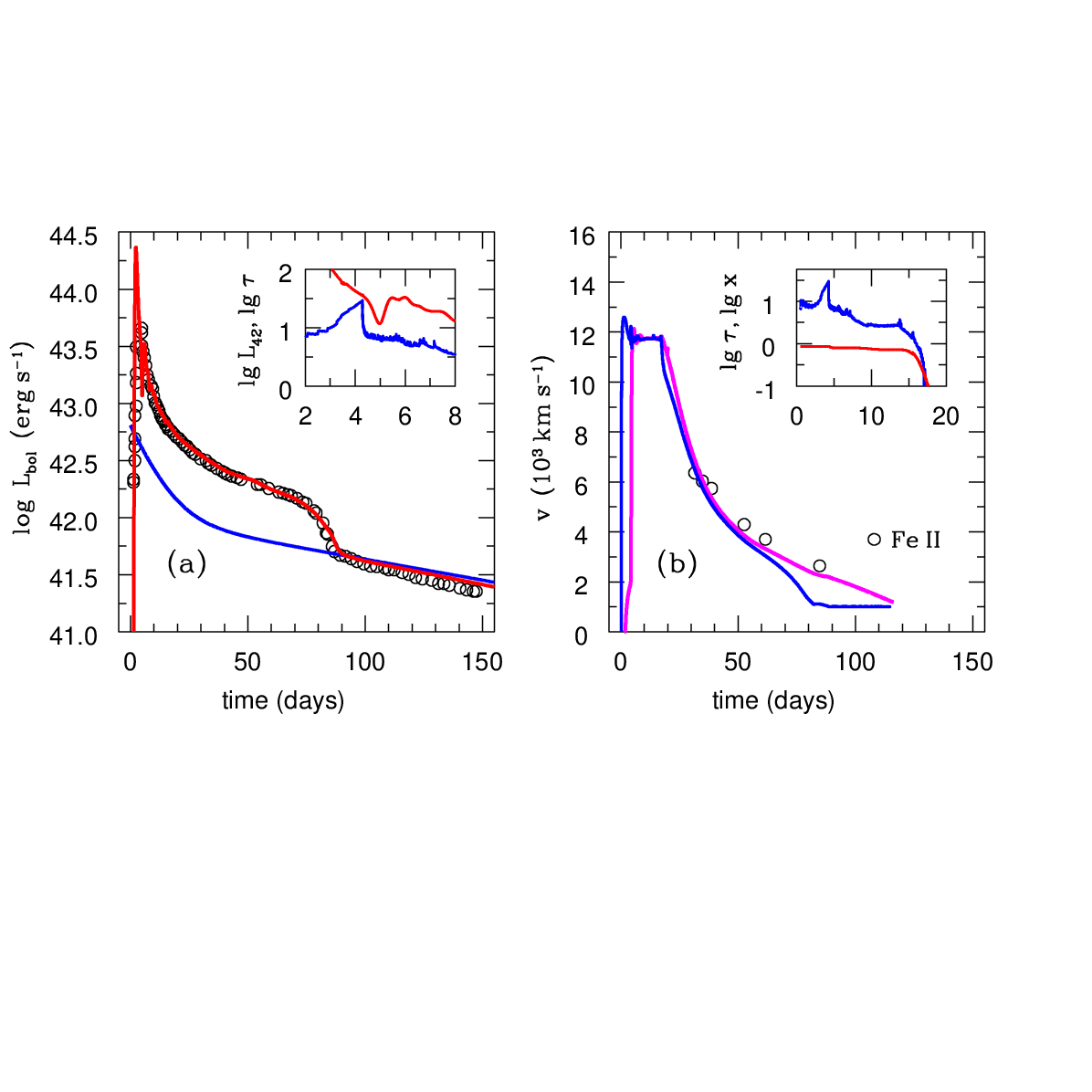}
   \caption{%
   The bolometric light curve and the evolution of photospheric velocity.
   Panel (a): the model light curve (\emph{red line\/}) overlaid on
      the bolometric data (\emph{circles\/}) \citep{Hsu_2025}.
   The \emph{blue line} is the total power of radioactive $^{56}$Ni decay.
   Inset shows the light curve fragment with a zigzag between days 5 and 7
      (\emph{red line\/}) and the CDS optical depth (\emph{blue line\/}).
   Panel (b): the evolution of model velocity defined by the level
      $\tau_{eff} = 2/3$ (\emph{blue line\/}) and $\tau_\mathrm{Thomson} = 1$
      (\emph{magenta line\/}) is compared with the photospheric velocities
      estimated from the absorption minimum of \FeII 5169\A \citep{Zheng_2025}.
   Inset shows the evolution of the CDS optical depth (\emph{blue line\/})
      and the hydrogen ionization fraction in the CDS (\emph{red line\/}).
   }
   \label{fig:lcv}
\end{figure*}
The optimal hydrodynamic model well reproduces both the bolometric light curve
   and the expansion velocities (Fig.~\ref{fig:lcv}).
Its basic parameters are the ejecta mass $M_{ej} = 13.2$\Msun, the explosion
   energy $E = 2.8\times10^{51}$\,erg, and the pre-SN radius $R_0 = 1540$\Rsun
   (Table~\ref{tab:param}).
Given the $^{56}$Ni mixing in velocity space up to 4000\kms, the $^{56}$Ni mass
   of 0.07\Msun is directly recovered from the radioactive tail.
The uncertainty in the derived SN parameters can be estimated by a variation
   of the model parameters around the optimal model \citep{Utrobin_2007}.
The uncertainties in the distance of $\pm0.14$\,Mpc \citep{Hsu_2025} and
   a total reddening of $\pm0.011$\,mag \citep{Singh_2024}
   imply the 5 per cent uncertainty in the bolometric luminosity.
The scatter in the plot of the photospheric velocity versus time around
   the middle of the plateau (Fig.~\ref{fig:lcv}b) suggests the uncertainty
   of 3 per cent in the photospheric velocity.
We estimate the maximal uncertainty of the plateau length as 2 days, i.e.,
   3 per cent of the plateau duration. 
With these uncertainties of observables, we find the errors of
   $\pm150$\Rsun for the initial radius, $\pm1.3$\Msun for the ejecta
   mass, $\pm0.4\times10^{51}$\,erg for the explosion energy, and
   $\pm0.004$\Msun for the total $^{56}$Ni mass (Table~\ref{tab:param}).

\begin{table}
\centering 
\caption{Parameters of the optimal model}
\label{tab:param}
\begin{tabular}{@{ } l @{ } c @{ } c @{ } c @{ }}
\toprule
\noalign{\smallskip}
Parameter & Unit & Value & Error \\
\noalign{\smallskip}
\midrule
\noalign{\smallskip}
Pre-SN radius              & \Rsun	         & 1540  & $\pm$\,150    \\
Ejected mass               & \Msun          & 13.2  & $\pm$\,1.3    \\
Explosion energy           & $10^{51}$\,erg & 2.78  & $\pm$\,0.4    \\
$^{56}$Ni mass             & \Msun          & 0.07  & $\pm$\,0.004  \\
Extent of $^{56}$Ni mixing & \kms           & 4000  & $\pm$\,300    \\
CS shell                   & \Msun          & 0.01  & --            \\
\botrule
\end{tabular}
\end{table}
The model light curve demonstrates an interesting feature --- a brief luminosity
   minimum around day 5 (Fig.~\ref{fig:lcv}a, inset).
The physics of this phenomenon involves three factors: the opaque CDS,
   the temperature decrease with time, and the specific temperature dependence
   of the Rosseland opacity: a sharp maximum at $10000-13000$\,K for
   the CDS density of $10^{-11}-10^{-10}$\gcmq.
The opacity maximum results in the increase of the CDS optical depth
   (Fig.~\ref{fig:lcv}a, inset), which in turn causes a brief lockup of
   the radiation flux followed by a subsequent escape of trapped radiation.
Note that the bolometric UVOIR ($0.16-2.35$\,$\mu$m) light curve
    seems to show a similar feature around day +5 \citep{Singh_2024}.

At the radioactive tail, after day 100, both the total power of radioactive
   $^{56}$Ni decay and the model bolometric luminosity are slightly higher
   compared to the observational light curve (Fig.~\ref{fig:lcv}a).
The model could produce a better fit with a larger extent of $^{56}$Ni
   mixing, but there is a tight observational constrain on the outer velocity
   of the $^{56}$Ni ejecta of 4000\kms from the high-quality
   [Co\,{\sc ii}] 10.521\,$\mu$m emission line \citep{Medler_2025}.
We leave the mentioned disparity unresolved, but notice that about 60\% of
   the bolometric luminosity at the radioactive tail is due to near- and
   mid-infrared radiation \citep{Jacobson_2025}.
As to the rate of gamma-ray deposition, we find that the gamma-ray escape
   fraction in our model is equal to $e^{-1} = 0.368...$ at the time
   $t_1 = 299.3$ days, in agreement with the observational estimate
   $t_1 = 313$ days reported by \cite{Jacobson_2025}.

The evolution of the photospheric velocity (Fig.~\ref{fig:lcv}b) reveals
   an interesting physics.
The initial photosphere velocity peak of 12500\kms is related to the CDS
   formed by the outermost RSG matter during the SBO.
After the brief velocity maximum, the CDS decelerates sweeping up the CS shell
   into the CDS.
This is demonstrated by the velocity evolution at the level
   $\tau_{\mbox{\tiny T}} = 1$ (Fig.~\ref{fig:lcv}b).
Later on the opaque CDS enters the rarefied RSG wind, where the CDS expands
   without deceleration that is indicated by the plateau of a constant velocity
   of about 11700\kms between days 6 and 18.
This behavior is consistent with the constant ($12000\pm500$\kms) radial
   velocity of the blue edge of Balmer line absorption (Sect.~\ref{sec:vel}).

The velocity plateau ends up with an abrupt drop of the photospheric velocity
   due to the rapid CDS clearing.
The latter is caused by the hydrogen recombination at around day 17,
   which is illustrated by the inset in Fig.~\ref{fig:lcv}b.
It shows that the CDS clearing coincides with the rapid decrease in the hydrogen
   ionization.
The latter signals that the declining CDS temperature reached the temperature
   of the hydrogen recombination ($\sim$8000\,K for the CDS density).

The SN~2023ixf ejecta mass combined with a neutron star of 1.6\Msun suggests
   the pre-SN mass of 14.8\Msun.
This indicates that the pre-SN could originate from a ZAMS progenitor with
   mass of about $16 - 17$\Msun, in accordance with a progenitor mass estimate
   of $17\pm4$\Msun based on the pre-explosion photometric data
   \citep{Jencson_2023}.
Somewhat surprising is the high explosion energy of $2.8\times10^{51}$\,erg,
   probably beyond energetic possibilities of the neutrino-driven explosion
   \citep{Janka_2017}.

\subsection{Emergence of hard X-rays in radiation hydrodynamics}
\label{sec:xrays}
Early emergence of hard X-rays \citep{Grefenstette_2023} provides us with an
   interesting insight into the transition from
   the radiation-dominated forward shock to the adiabatic high-temperature
   matter-dominated shock.
This transition cannot be reproduced in the picture of the   
   interaction of the cold SN ejecta with the undisturbed CSM, likewise in the 
   SN 1993J model for hard X-rays \citep{Suzuki_1995}.
    
The point is that after the SBO the powerful SN radiation accelerates
   the preshock CS gas up to velocities comparable to the shock speed
   \citep{UC_2025}.
As a result, a viscous jump becomes very weak and the post-shock gas temperature
   remains comparable to that of the SN radiation --- the feature of
   the radiation-dominated shock \citep{ZR_1967}.
While the forward shock propagates, the preshock velocity decreases,
   the viscous jump becomes larger, and at some moment the post-shock gas
   temperature sharply rises above $10^8$\,K.
The evolution of hydrodynamic values (velocity, density, and temperature)
   during the transition from the radiation-dominated to matter-dominated regime
   has been described in detail for SN~2024bch \citep{UC_2025};
   the overall evolution in the case of SN~2023ixf is very similar.

\begin{figure}
   \includegraphics[width=\columnwidth, clip, trim=9 239  49 121]{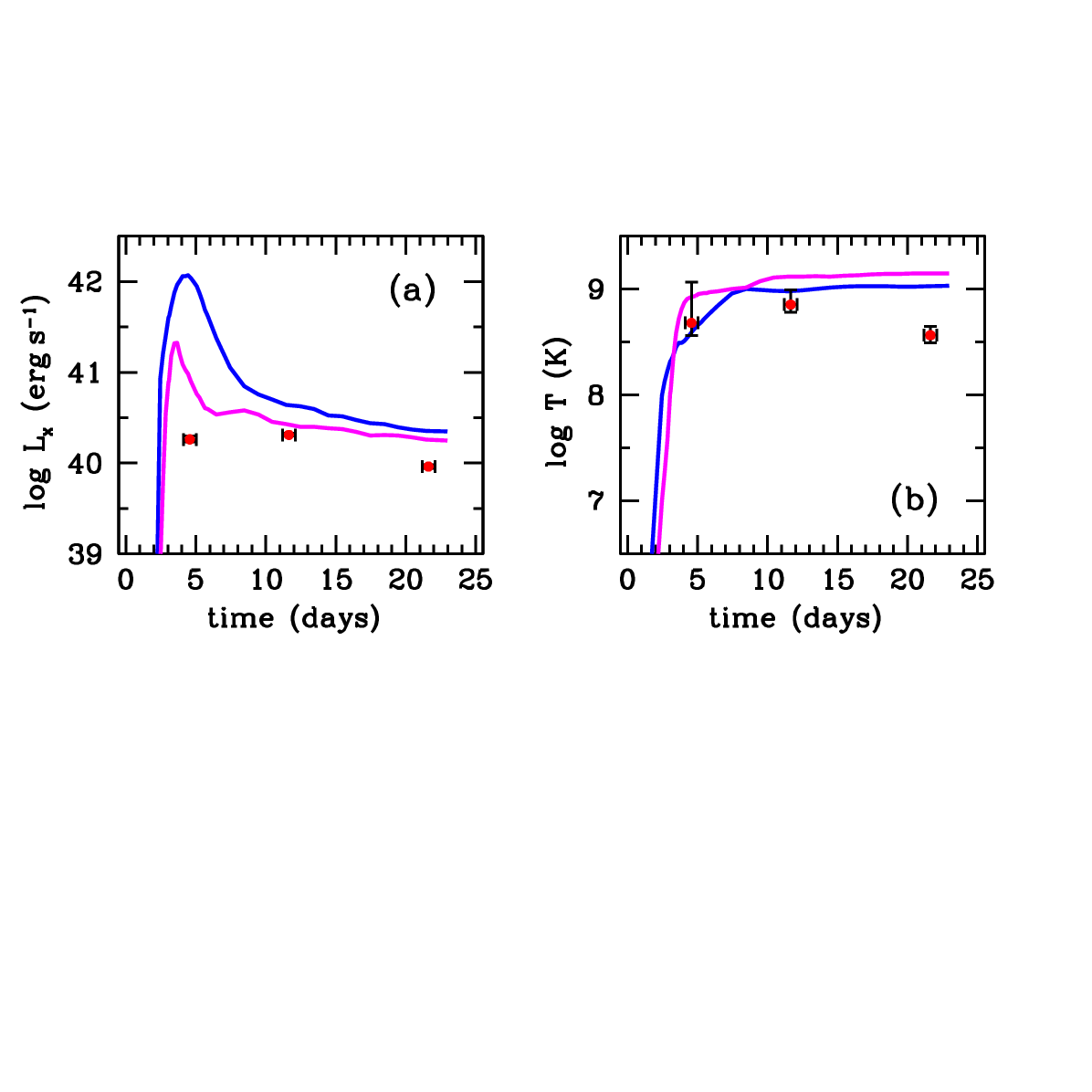}
   \caption{%
   The emergence of hard X-rays in SN~2023ixf.
   Panel (a): the early evolution of the hard X-ray luminosity in the frame of
      radiation hydrodynamics in the case of the CSM density inferred
      from the \Ha line profile and the X-ray column density
      (see the inset in Fig.~\ref{fig:csha}b) (\emph{blue line\/})
      and in the case of 10 times lower density of the CS shell
      (\emph{red line\/}).
   \emph{Red} circles are the luminosity values obtained from X-ray data
      \citep{Nayana_2025}.
   Panel (b): the electron temperature of the shock emitting X-rays.
   See Panel (a) legend for details.
   }
   \label{fig:xray}
\end{figure}

The radiation hydrodynamic model of SN~2023ixf results in the early evolution
   of the hard X-ray luminosity and electron temperature (Fig.~\ref{fig:xray})
   that are in satisfactory agreement with the {\em NuSTAR} data
   \citep{Nayana_2025}.
We adopt equal post-shock electron and ion temperatures ($T_e = T_i$).
This approximation is justified at the early ($t < 10$ days) stage
   \citep{UC_2025} and responsible for the larger $T_e$ value compared to
   the observational estimate on day 22 (Fig.~\ref{fig:xray}).
The model with the CS density suggested by the Thomson optical depth
   inferred from the \Ha line and the column density inferred from X-ray data
   (Fig.~\ref{fig:csha}b, inset) predicts a higher X-ray luminosity compared to
   the observational value on day 5.
A better agreement is found for the model with 10 times lower density of
   the CS shell.
This fact can be interpreted as an evidence for the clumpy structure of
   the CS shell in which case the intercloud rarefied medium is responsible
   for the hard X-ray emission, whereas the clumpy dense component is
   responsible for the X-ray absorption.

\subsection{Clumpiness of CS shell}
\label{sec:clump}
%
The suggested clumpy structure of the CS shell should meet several requirements.
First, the filling factor of the intercloud medium must be close to unity
   and the intercloud density $\rho_{ic}$ must be low compared to
   the average density $\rho$,  $\rho_{ic} = \eta \rho$ with $\eta \approx 0.1$.
The latter means that the intercloud component does not contribute to
   the column density inferred from X-ray data on day 5.
Second, the clumpy component has to provide a full occultation of
   the X-ray-emitting shock in order to be compatible with the column density
   inferred from X-ray data on day 5.

The full occultation of the shock wave suggests that a mean number of clouds
   (an occultation optical depth $\tau_{oc}$) along the length comparable to
   the radius of the CS shell $R_{cs}$ is greater than unity.
With the filling factor of cloudy component $f_c$ and the average clump
   radius $a$, a total length of clump secants along the line segment of
   length $R_{cs}$ is of $f_c R_{cs}$ \citep{Kendall_1963}.
Given a mean secant of a spherical cloud of $4a/3$, the occultation optical
   depth is $\tau_{oc}  = (f_cR_{cs})/(4a/3)$.
The condition for the full occultation $\tau_{oc} \gtrsim 1$, therefore,
   imposes an upper limit on the clump radius $a \lesssim f_cR_{cs}$.

The third requirement is that clumps should not affect the intercloud medium
   during the shock wave propagation in the CS shell.
This condition means that a fragmentation and dispersal of shocked clouds
   occur on a timescale larger than a crossing time of the CS shell by
   the forward shock $t_{cross} \approx R_{cs}/v_{sh}$.

The density contrast of clouds vs. the intercloud medium is
\begin{equation}
\chi = \frac{\rho_c}{\rho_{ic}} = \frac{1-\eta}{\eta f_c}\,
\end{equation}
For $\eta = 0.1$ and a fiducial filling factor $f_c = 0.01$ the density contrast
   is $\chi = 900$.
A stripping of a shocked cloud occurs on the timescale \citep{Klein_1994}
\begin{equation}
	t_{st} \approx Ft_{cc} = F\frac{a\chi^{1/2}}{v_{sh}}\,,
	\label{eq:strip}
\end{equation}
   where $t_{cc}$ is the cloud crushing time, $v_{sh} = 12000$\kms, and
   the factor $F > 1$.
The two-dimensional hydrodynamic simulations of the shock wave interaction
   with dense cloud ($\chi \gtrsim 100$) show that it takes $\approx$7$t_{cc}$
   to strip away about 75\% of a shocked cloud \citep{Klein_1994}.
Experiments with shock tube at the Omega laser facility produce similar result:
   80\% of shocked cloud is stripped for $\approx$8$t_{cc}$ \citep{Hansen_2007}.

Adopting $F = 8$, $a \approx f_cR_{cs}$, $f_c = 0.01$, and $\chi = 900$,
   we find from Equation (\ref{eq:strip}) $t_{st} \approx 2.4t_{cross}$.
This means that for sensible parameters the forward shock propagation in
   the intercloud medium of the CS shell is almost not affected by the shock
   interaction with the cloudy component.

\section{Summary}
\label{sec:sumry}
The goal of the paper has been to obtain the key parameters of SN~2023ixf
   from the hydrodynamic modeling based on the essentially broader set of
   observational data compared to the previously released models.
We recover the early-time maximum expansion velocities from the shallow Balmer
   absorption lines \citep{Zheng_2025}, which is a decisive factor in removing
   the parameter degeneracy.

We infer the explosion energy of $2.8\times10^{51}$\,erg, the ejecta mass
   of 13.2\Msun, the pre-SN radius of 1540\Rsun, and the $^{56}$Ni mass of
   0.07\Msun.
The pre-SN model is supplemented by the CS shell (internal wind) with
   the mass of 0.01\Msun and the wind density parameter
   $w_{int} = 3.7\times10^{16}$\gcm.
The external wind has about 20 times lower density parameter
   $w_{ext} = 1.7\times10^{15}$\gcm.
   
Based on the radiation hydrodynamic model, we simulate the SN~2023ixf explosion
   in the dense CS environment that is accompanied with the emergence of
   hard X-rays on day 5 in agreement with the observations provided the clumpy
   structure of the CS shell.

To our knowledge, the transition of the SN/CSM interaction from the
   radiation-dominated SBO to the almost adiabatic matter-dominated regime
   with the subsequent emergence of hard X-rays from the forward shock
   has never been described for SNe in the framework of the radiation
   hydrodynamics.

\backmatter

\bmhead{Acknowledgements}
Not applicable.

\bmhead{Author contribution}
The authors contributed equally to this work.

\bmhead{Funding}
Not applicable.

\bmhead{Data availability}
No datasets were generated or analyzed during the current study.

\bmhead{Materials availability}
Not applicable.

\bmhead{Code availability} 
Not applicable.

\section*{Declarations}

\subsection*{Ethics approval and consent to participate}
Not applicable.

\subsection*{Consent for publication}
Not applicable.

\subsection*{Competing interests}
The authors declare no competing interests.


\end{document}